\documentclass[journal, 10pt]{IEEEtran}

\ifCLASSINFOpdf
  \usepackage[pdftex]{graphicx}
\else
\fi

\usepackage{amsmath}
\usepackage{multirow}
\usepackage{color}

\usepackage{epsfig} % for postscript graphics files

\usepackage{xcolor}

\ifCLASSOPTIONcompsoc
  \usepackage[caption=false,font=normalsize,labelfont=sf,textfont=sf]{subfig}
\else
  \usepackage[caption=false,font=footnotesize]{subfig}
\fi
\usepackage{url}
\usepackage{balance}
\usepackage[inline]{enumitem}
\usepackage[normalem]{ulem}
\usepackage{tabularx,booktabs}
\newcolumntype{C}{>{\centering\arraybackslash}X} % centered version of "X" type
\usepackage{lipsum}
\usepackage{booktabs}% http://ctan.org/pkg/booktabs

\usepackage{xcolor,colortbl}
\usepackage{adjustbox}
\usepackage{hhline}
\usepackage{soul}
\usepackage{makecell}
\usepackage{hyperref} 
\usepackage{tabularray}

\begin{document}
\title{
Consolidate Viability and Information Theories for Task-Oriented Communications: \\A Homeostasis Solution
}

\author{Ozgur Ercetin, Mohaned Chraiti and Rustu Erciyes Karakaya \thanks{The authors are with 
the Faculty of Engineering and Natural Sciences, 
Sabanci University, Istanbul Turkiye. Emails: \{oercetin, mohaned.chraiti, rerciyes\}@sabanciuniv.edu This work is supported in part by The Scientific and Technological Research
Council (TUBITAK) of Turkiye under grant number 119E353.}}
%e-mail: oercetin@sabanciuniv.edu

\addtolength{\oddsidemargin}{-1mm}
\addtolength{\textheight}{1mm}
\setlength{\baselineskip}{10pt}
\setlength{\columnwidth}{241pt}

\maketitle
%\renewcommand{\baselinestretch}{0.87}

\begin{abstract}

  The next generation of cellular networks, 6G, { \color{black}is expected to offer a range of exciting applications and services, including holographic communication, machine-to-machine communication, and data sensing from millions of devices. There is an incremental exhaustion of the spectral resources. It is crucial to efficiently manage these resources through value-driven approaches that eliminate waste and continually enhance the communication process. These management principles align with the Task-Oriented Communications (TOC) philosophy. The aim is to allocate the minimum necessary communication resource according to the receiver's objective and continuously improve the communication process. However, it is currently unclear how to build knowledge of the receiver's goal and operate accordingly for efficient-resource utilization. Our management approach combines viability theory and transfer entropy to ensure that the actor remains within a viable space as per their goal and to gradually reduce the information exchange through knowledge accumulation. We discuss these theories in the context of TOC and examine their application in the plant process control case. Finally, we provide insights into future research directions from computational, performance, and protocol perspectives.}
\end{abstract}

%\renewcommand{\baselinestretch}{1}

%==========================================

\section{Introduction}

{\color{black} Shannon's influential work, which falls under the category of Level A Technical Communications, has provided a strong foundation for groundbreaking advancements in communications and networking. The objective of Technical Communications is to ensure that a sequence of bits is perfectly reconstructed at the receiver. This approach was effective in an era when end devices had limited computational capabilities. However, in the last five decades, the computational capabilities of end devices have grown exponentially. Additionally, the use of specialized artificial intelligence (AI) chips in end-user devices has become more common, enabling the application of modern machine learning algorithms for highly complex calculations. In this respect, discussions have already begun on a paradigm shift from Shannon's Level-A to Level-B (semantic)/Level-C (effective) of communications \cite{strinati20216g}. Some of the recent works approach the problem from a semantic perspective, i.e., focusing on the conveyed meaning of information. The objective is to reduce the number of bits in transfer by performing {\em end-to-end} semantic encoding \cite{strinati20216g}. However, by itself, semantic communications do not capture the time-dependent impact on the receiver.  Recent works are increasingly directing their focus toward task-oriented communications, in which the impact of information on the receiver is central to the investigation. In \cite{10122224}, the authors explore the role of fidelity in goal-oriented semantic communication through a rate-distortion approach. Another study proposes an explainable semantic communication that selectively transmits task-relevant features for improved transmission efficiency and robustness against semantic noise \cite{10110357}. In another work, the authors investigate the possible cooperation between the senders and receivers to minimize semantic error (i.e., belief efficiency).and achieve a goal via curriculum learning \cite{curriculumLearning}.}

{\color{black} This paper discusses the design of task-oriented communications (TOC) pertaining to \textit{the afterword impact of the received data on the end-user's actions/behaviors/feelings}. There is no doubt that the receiver must properly decode/interpret the received data before taking a possible action; however, the effect of this interpretation on the receiver's actions differentiates TOC from semantic communication. The main utility of TOC is to continuously assess the need/goal of the receiver and its capability, then accordingly assign the minimally sufficient communication resource. For the majority of case scenarios, \textbf{the receivers (actors) goals outlines are flexible to a certain extent, and \textit{robust}, their actions do not suffer from limited irregularities.}  The receiver's robustness results from the processes’ characteristics or the AI-empowered knowledge base that takes action. Whether or not the actor has strict goal outlines or is robust to irregularities, these quantities have to be understood, analyzed, and exploited in an autonomous (closed-loop) fashion for a communications approach that delivers values as per the actor's object.}

{\color{black} TOC leverages continuous understanding/assessing of the actors' goals, capabilities, and environment to deliver adequate communications resources. In light of irregularities and unforeseen events, ranging from the actor's inability to execute control-unit directives with precision (internal impairment) to those caused randomly by nature (external impairment), the viability theory \cite{viability} provides an adequate framework for TOC. This theory defines the viable space, also known as the viability kernel, within which an actor can evolve toward its goals while preserving important qualities such as adaptation, stability, confinement, homeostasis, and tolerance. The larger the kernel size, the more tolerant the actor is to errors, including those resulting from delayed instructions, thereby reducing further resource utilization. }

{\color{black} Although the viable space can be derived through optimization or machine learning, it differs conceptually and in terms of its core objective. The viability theory offers more thorough solutions, as it assesses the risks of adopting a solution in the presence of uncertainty. In contrast, optimization approaches typically determine the best strategy, often overlooking close-performing solutions in which the actor remains operative. As a result, the actor may need to continuously seek support from the control unit, in case it is impelled outside the thin-line solution.}

{\color{black} The viability kernel encompasses a set of viable states, providing actors with room for action and decision errors, enabling more autonomous task-execution. The actor resorts to the control-unit only when events require more refined solutions, computationally demanding decisions, and/or more trained models. The viability kernel width determines the resilience degree to unknown disruptions. Understanding the viability kernel allows us to derive the appropriate communication rate required to convey minimum-sufficient directives and knowledge. As per the viability theory, the objective of TOC is to dynamically analyze the environment and actor state to develop strategies that maintain a safely-large viable space.}

{\color{black} For example, an efficient communications strategy may involve adopting a viable path with a low communication rate, instead of a shorter path that requires channeling high-rate instructions. The viable space depends largely on the actor's capability and knowledge, which can be continuously accumulated from the past. To measure the rate of meaningful information flow over time while knowledge continues to be accumulated, we suggest using Transfer Entropy (TE) \cite{transfer_entropy}. This measure quantifies the knowledge transfer between communicating parties, enabling them to assess and adjust their communication strategy.}

%\vspace{-9mm}
\section{Motivating Use Cases}
%In this section, we analyze two examples through which we discuss the viability theory for TOC.
%\vspace{-1mm}
\subsection{Multi-sensory and holographic telepresence}
%{\bf } --  %Let us 

{\color{black} Consider the case of replicating the sense of touch at a remote location \cite{9679801}. The receptors on the hand/feet from the brain are about 1.5 and 2 meters away, and the speed at which a sensory point transmits data is around 30 m/s. Accordingly, the maximum delays of signals are 0.038s and 0.067s \cite{9060950}. A human can detect temperature differences with a 0.02$^\circ$C resolution in $5-45^\circ$C range. Collecting information with a sufficient sampling rate of approximately 50 sample/s, the required bit rates for transmitting temperature and pressure information replicating the sense of touch can reach 880 Mbps with a maximum end-to-end delay of a few tens of milliseconds. However, the human brain can interpolate lost or irregular data based on prior experiences and only react to significant changes that deviate from the acceptable ranges. Thus, transmissions below peak speeds may not negatively affect a person's experience.}

{\color{black} Viability theory helps identify the set of communication bit-rates and delays sufficient to replicate a human sensing experience. Additionally, transfer entropy quantifies the information flow between different sensory points and assists in identifying information redundancies and dependencies. These tools ease the design of efficient TOC systems that cater to multi-sensory and holographic telepresence applications.}

 \subsection{Autonomous Vehicles}
 {\color{black} Consider the example of autonomous vehicles that not only aim to minimize travel time/trajectory but also prioritize safe navigation in the presence of obstacles and potential hazards. The vehicle employs trained models and gets support from a remote control-unit that is assumed to be more knowledgeable and more computationally capable. The AI-models at the control-unit result from the experience of multiple vehicles in different environments. To minimize the risk of safety events, a possible solution is to ensure continuous communication between the vehicle and a cloud-based control-unit, to get support in the event of an unforeseen safety issue that the vehicle model may not be equipped to handle on its own.}

{\color{black} A viability approach considers several factors, including the environment, vehicle capability, and control-unit support. The goal is to ensure that the vehicle can safely evolve between states while adhering to system constraints like delay and battery-life. The vehicle and control-unit adjust accordingly communication rates to maintain a large viable space. Higher communication rates are adopted when the vehicle is near the edge of the viable space. As the environment changes, the viability theory continually reassesses risks and system constraints, increasing vehicle autonomy and reducing communication rates as needed. This creates a task-oriented approach to communication. }

\subsection{The statement for task-oriented communications}

{\color{black} In the listed examples, the goal is not to support continuous streaming of timely packets from sensors or minimize the mismatch between the actual and estimated data. Instead, it is to design a system that functions similarly to a human communication/processing system, which is task-oriented and aligns with the viability theory principle. The primary aim is to allocate resources to communicate changes that fall outside the viable space to ensure a genuine and swift response from the system. For example, a person responding to significant changes in temperature/surface tension, a doctor responding to changes in a remote patient's body, or an autonomous factory responding to anomalies in machines' states. }

{\color{black} TOC solutions typically answer two fundamental questions: \textbf{1) How can we communicate effectively while ensuring proper operations of the end-system? 2) What is the optimal data rate given a task to accomplish?} Our TOC framework incorporates two key concepts to answer these questions: \textbf{1) The viability theory to identify and expand the actor's space of options to ensure robust operation. 2) The transfer entropy to measure knowledge accumulation and utility of communications while executing current and future tasks.} The following is an overview of these concepts.}

%\vspace{-0.2cm}
\section{Information and Control Theoretic Tools for Task-oriented Communications}

\subsection{Viability Theory}
%%%%%%%%%%%%%%%%%%%%%%%%%%%%%%%
%  THE FOLLOWING IS THE TEXT FOR MAGAZINE
 \begin{figure}
     \centering
     \includegraphics[width=2.8in]{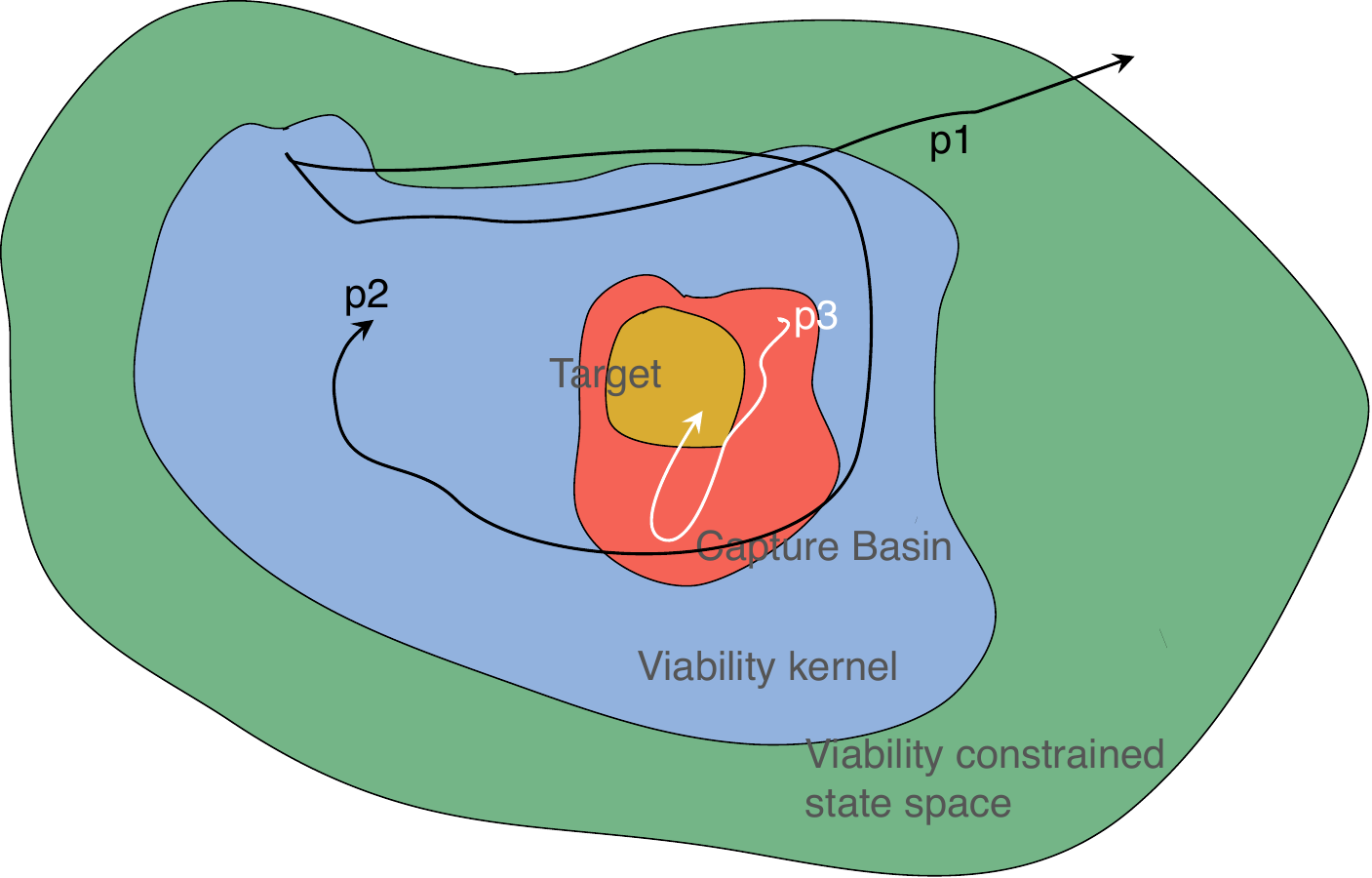}
     \caption{Viable control.}
     \label{fig:viability}
 \end{figure}
 
 {\color{black} Viability theory is a field of mathematics that explores knowledge on dynamic systems evolution in the presence of state constraints and uncertainties \cite{viability}. Then, it produces a set of solutions or states known as the ``viability kernel." The latter represents the set of initial states from which the actor can start and maintain viability for the duration of its operation. Operating at the boundary of the viability kernel enables the actor to prioritize critical information transmission. In contrast, an expanded viability kernel indicates a higher degree of tolerability to errors and uncertainties, allowing for adjusted communication rates and conserving resources. By leveraging an understanding of the boundaries of the viability kernel, the actor can optimize its communication strategy to adapt to different states within the viability kernel, enhancing the efficiency and resilience of the system.}

{\color{black} The main concepts of viability theory are depicted in Figure \ref{fig:viability}. The set of states in which an actor can operate safely is known as the viability kernel, which is denoted as $\operatorname{Viab}(g, U, M)$. It represents the initial states from which there exist controls and trajectories that satisfy the constraints for all positive times. However, not all trajectories starting from the viability kernel are admissible, as illustrated in Figure \ref{fig:viability}. To address time and delay constraints, the concept of a capture basin shall be used. The capture basin represents the set of viable current states that can achieve a specific task within a predefined time window. The Hamiltonian function is commonly used to analyze a system's viability. The function combines the state variables, co-state variables, and control variables. The co-state variables are associated with a normal cone at a particular state point, providing information on the feasible directions in which the actor can move while still satisfying the constraints. The viability condition requires that for each possible direction at a given state, there exists a control such that the Hamiltonian function at that point is non-positive.}

{\color{black} To put it simply, the viability-based approach to TOC requires identifying viable spaces and paths, while considering constraints, performance targets, uncertainties, and actor capabilities. The system then adjusts communication resources in a closed-loop to maintain viability. The size of the viable space depends on the actor's capabilities, which can be enhanced to improve the viable space and optimize communication resource usage.
}

\vspace{-3mm}
\subsection{Transfer Entropy}
%%%%%%%%%%%%%%%%%%%%%%%%
% THE FOLLOWING IS THE TEXT FOR MAGAZINE.

{\color{black}Transfer entropy (TE) is a measure used to quantify the flow of information between random processes by analyzing causality and deviations from the generalized Markov property \cite{transfer_entropy}. Transfer entropy evaluates the impact of past values of the control-unit ($Y$) on the current value of the actor's action ($X$). This metric is an upper limit of another information-theoretical measure called directed information, indicating improved communication effectiveness with accumulated contextual knowledge. It can be expressed as the sum of conditional mutual information terms and characterizes the overall stochastic system dynamics and control data. }

{\color{black} By integrating transfer entropy with viability theory, we can measure the transfer of knowledge between past and current/future actions or events. The actor-centric operation allows for customization based on diverse end-user perceptions and applications. The actor's behavior relies on accumulated and inherited knowledge, and the upper limit of this knowledge can be represented by the information content derived from the history of two random processes: the control-unit control action random process and the actor state random process.}

\section{Proposed Framework}

\begin{figure}
     \centering
     \includegraphics[width=2.3in]{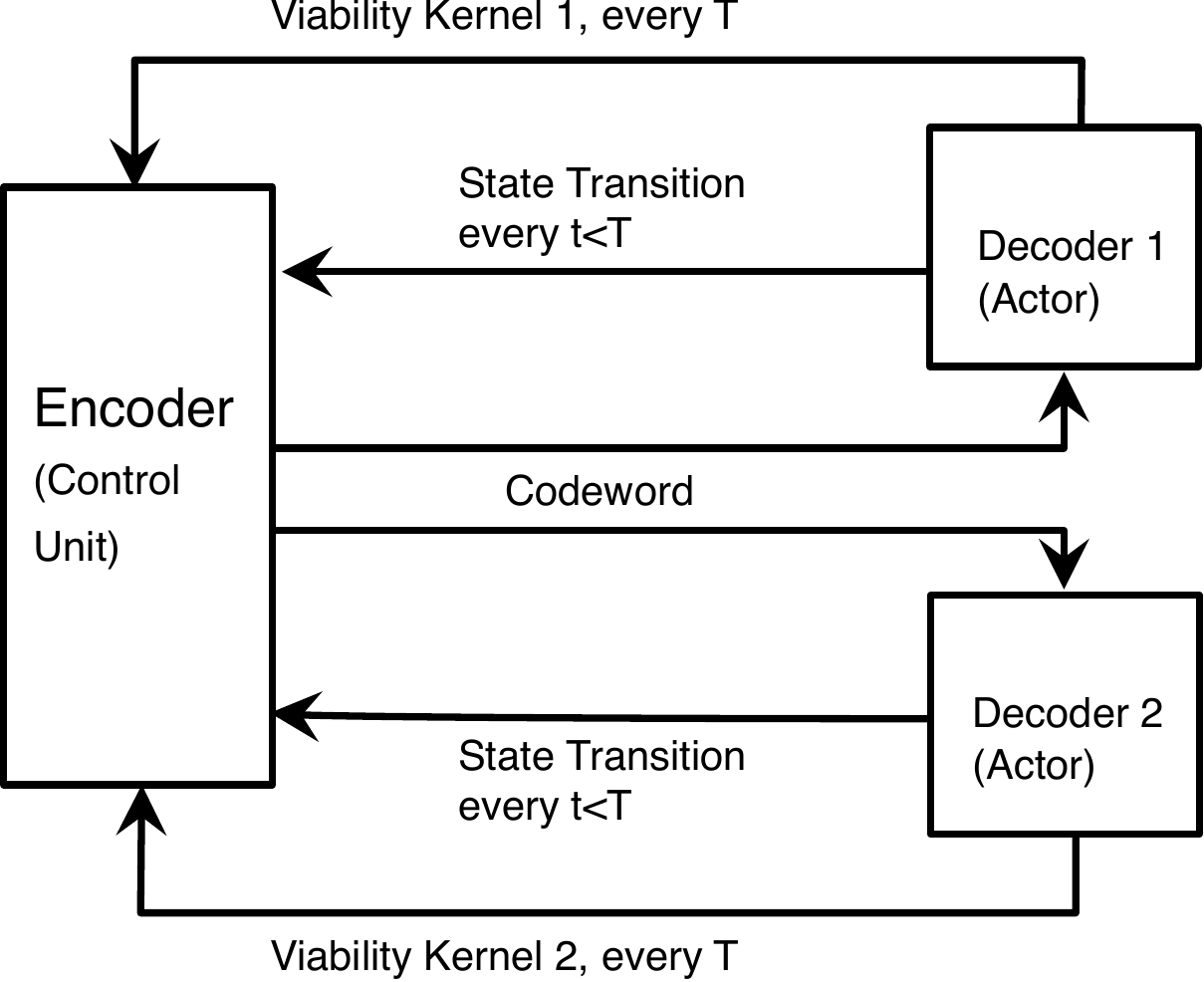}
     \caption{Proposed framework to ensure safe and successful operation when a control unit communicates with two different actors. As knowledge is transferred, we anticipate that actors will require fewer updates on their state transitions, and hence, the control unit will provide fewer control directives.}
     \label{fig:framework}
 \end{figure}

{\color{black} Viability theory is a framework for examining possible states and identifying those that are ``viable", as well as the transitions between them. These states can include adaptation, stability, confinement, homeostasis, tolerance, and more. The choice of criteria depends on the actor (or ``agent"), their capabilities, and the environment. These factors can change over time as the actor's situation evolves. In TOC, the viability kernel is discussed in the context of an actor seeking to achieve its objective with the help of a remote decision unit that provides minimal-necessary support. The control unit starts out more knowledgeable than the actor, with trained models, experience, and statistics. However, the environment and actor responses can be unpredictable, leading to unexpected events like glitches and disturbances. See Fig.~\ref{fig:framework} for the system model.}

{\color{black} The actor and control unit are continually communicating in a closed-loop to achieve two goals: dynamically identifying and adjusting the viable kernel, and accordingly the appropriate communication rate. The size and characteristics of the viable kernel largely govern the rate of information exchange. When the viability kernel is broad and/or the current state is far from the boundary, the actor will not steer into a non-viable state with high confidence. The actor will be able to tolerate control decision noise and delayed instructions, and a low-rate update of control instructions will suffice. This means that the actor will have more autonomy in decision-making, without requiring pinpoint accuracy to restore viability. As a result, low information exchange rates and reliable communication will be ensured between both parties.}

{\color{black} Viability theory offers adaptivity in terms of metrics. When stringent constraints are involved, it is more suitable to consider the capture basin instead of the viability kernel. This allows for instance a drone with delay constraints to consider the fastest path, even if it involves more communication resources. Among a myriad of choices, it is important to establish a framework for the decision policies. Besides, the viable space heavily depends on the capability of the actor. Therefore, it is not to underestimate the capacity of the actor to learn from past decisions to enhance its knowledge and build decision-making intelligence, i.e., transfer of knowledge. Present knowledge and the history of evolution help in decision-making through myopic selection mechanisms instead of inter-temporal optimization. The actor becomes with time more autonomous, reducing the need for communication. This is where directed information and transfer entropy come into play as the second part of the framework, quantifying the transfer of knowledge between the control-unit decisions random process, actor state random process, and probabilistic model that binds both processes.  Quantifying the transfer of knowledge will be the key to gradually reducing the exchange rate of information toward efficient TOC.}

{\color{black}The proposed TOC system setup is depicted in Figure~\ref{fig:framework}. In order to ensure the continuous operation of the system, it is imperative to take into account the bidirectional flow of information between the random actor process and the control process. The encoder and decoder can reduce transmission rates with the consultation of the viability kernel. If the viability kernel is broad or the current state is distant from the boundary, the likelihood of the system becoming non-viable is small. Additionally, if the decoder does not heavily rely on the encoder for viable actions, the codeword and state transition flow rates can be minimized. It is essential to note that the viability kernel is dynamic due to the control policy evolving with accumulated knowledge.}

%\vspace{-5mm}
\section{An Illustrative Case Study}
{\color{black} We demonstrate the distinguishing features of viability theoretical TOC policies via a dynamic control example taken from \cite{10.1007/3-540-60472-3_7}.  We consider the case of the plant (actor) remotely operated by a control-unit similar to the case of the Factory-of-the-Future (FoF). The plant has the capability of sensing the data and communicating them to the control-unit and adjusting the system functioning according to the control-unit directives with reasonable precision, but not free of error. The plant dynamics are characterized by nonlinear differential equations. The process requires cycling the plant repeatedly through three operating points. The process is \textit{robust} so that it is sufficient to visit some specified neighborhoods of these operating points. The plant is described with state variables; the temperature of the plant and its pressure. Their units of measurement are normalized so that $(0, 0)$ represents ambient conditions. There are two control inputs, i.e., the rate at which heat is supplied to the plant, and the rate at which a pneumatic piston is displaced.}

\begin{figure*}[!t]
    \centering
    \begin{tabular}[c]{cc}
        \subfloat[Fixed update \label{a}]{%
        \includegraphics[width=0.45\linewidth]{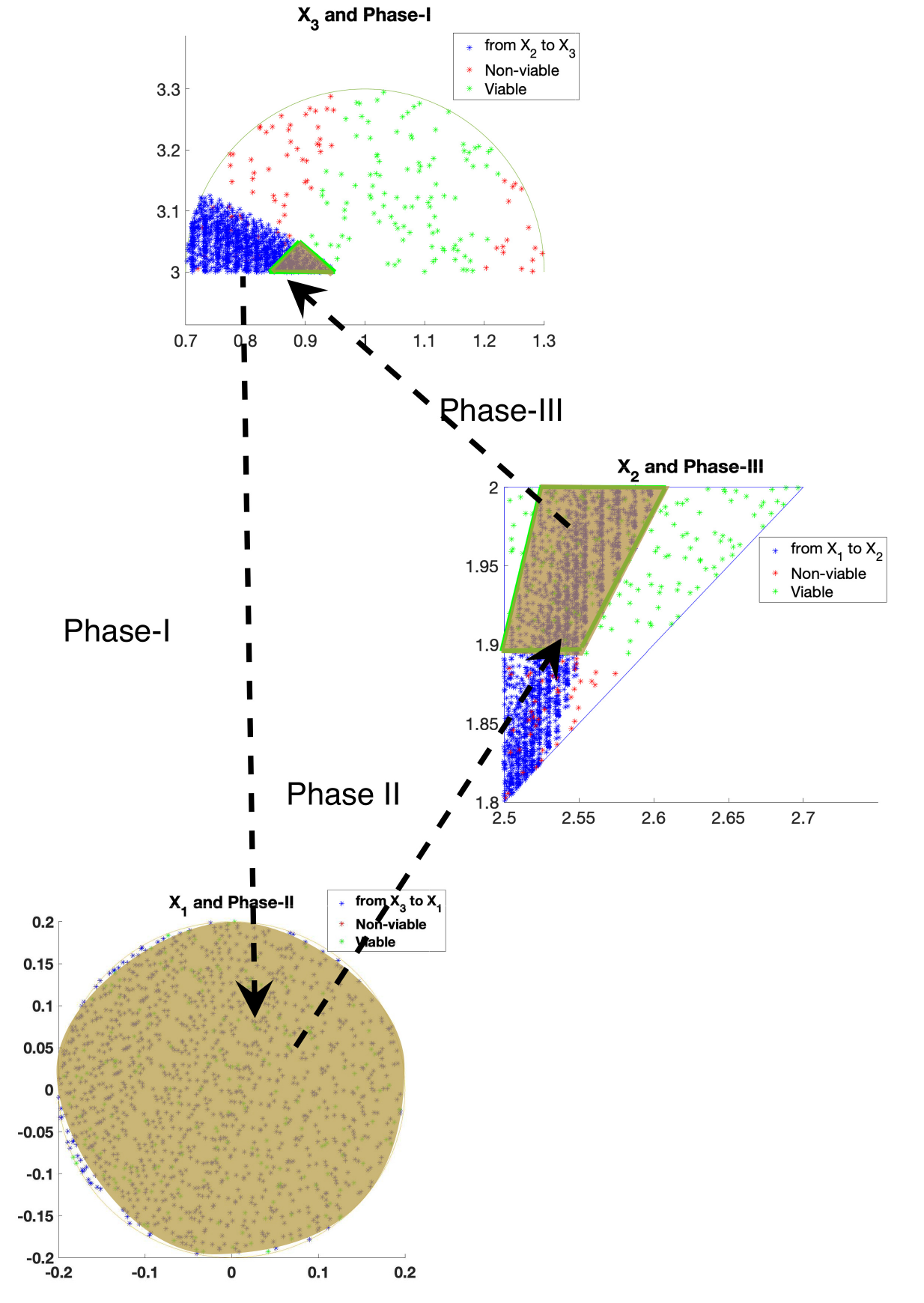}} 
        &
        \begin{tabular}[b]{c}
             \subfloat[Viable points in $X_3$ with adaptive update. \label{b}]{%
        \includegraphics[width=0.35\linewidth]{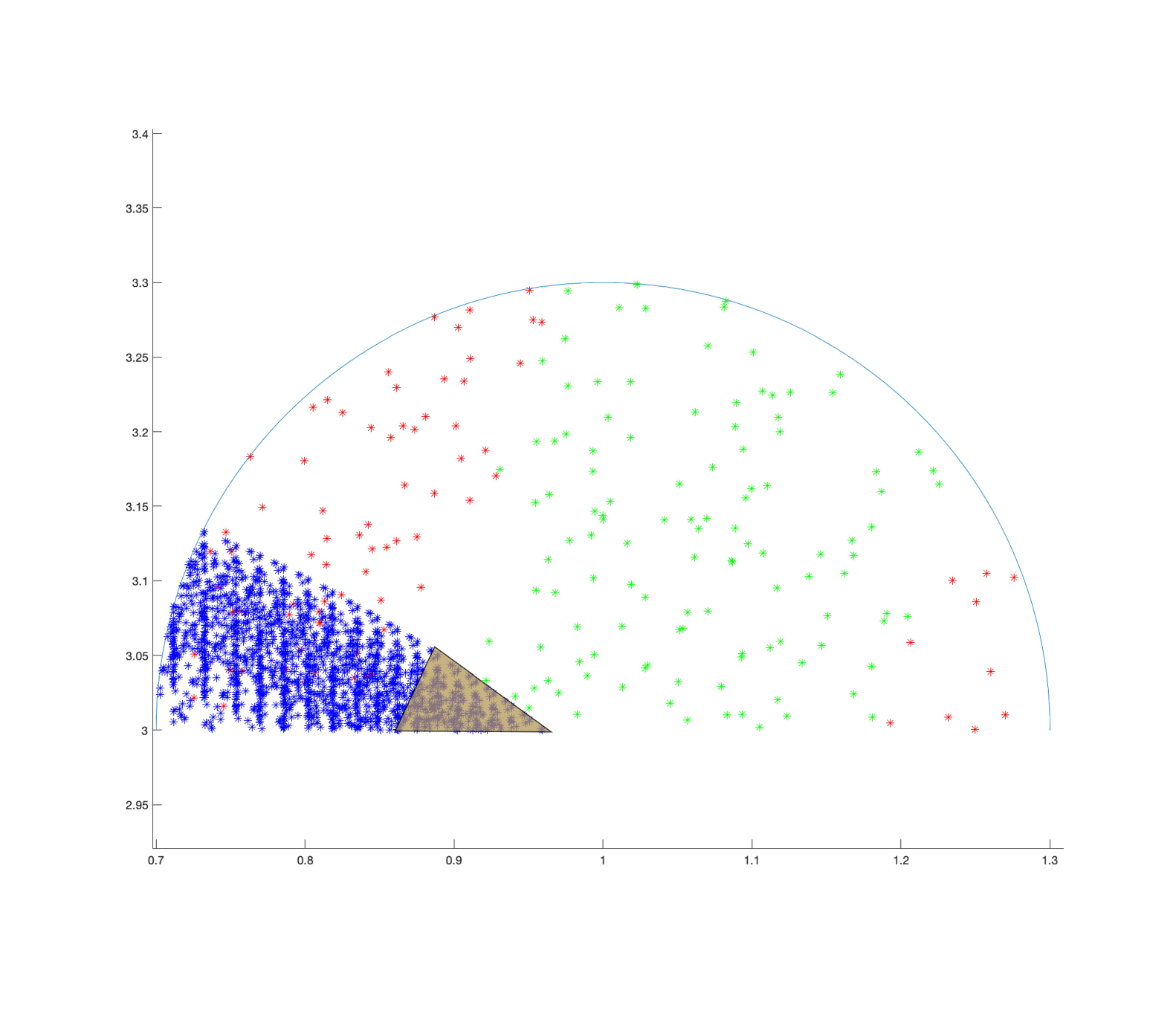}}  \\
             \subfloat[Viable points in $X_2$ with adaptive update. \label{c}]{%
        \includegraphics[width=0.35\linewidth]{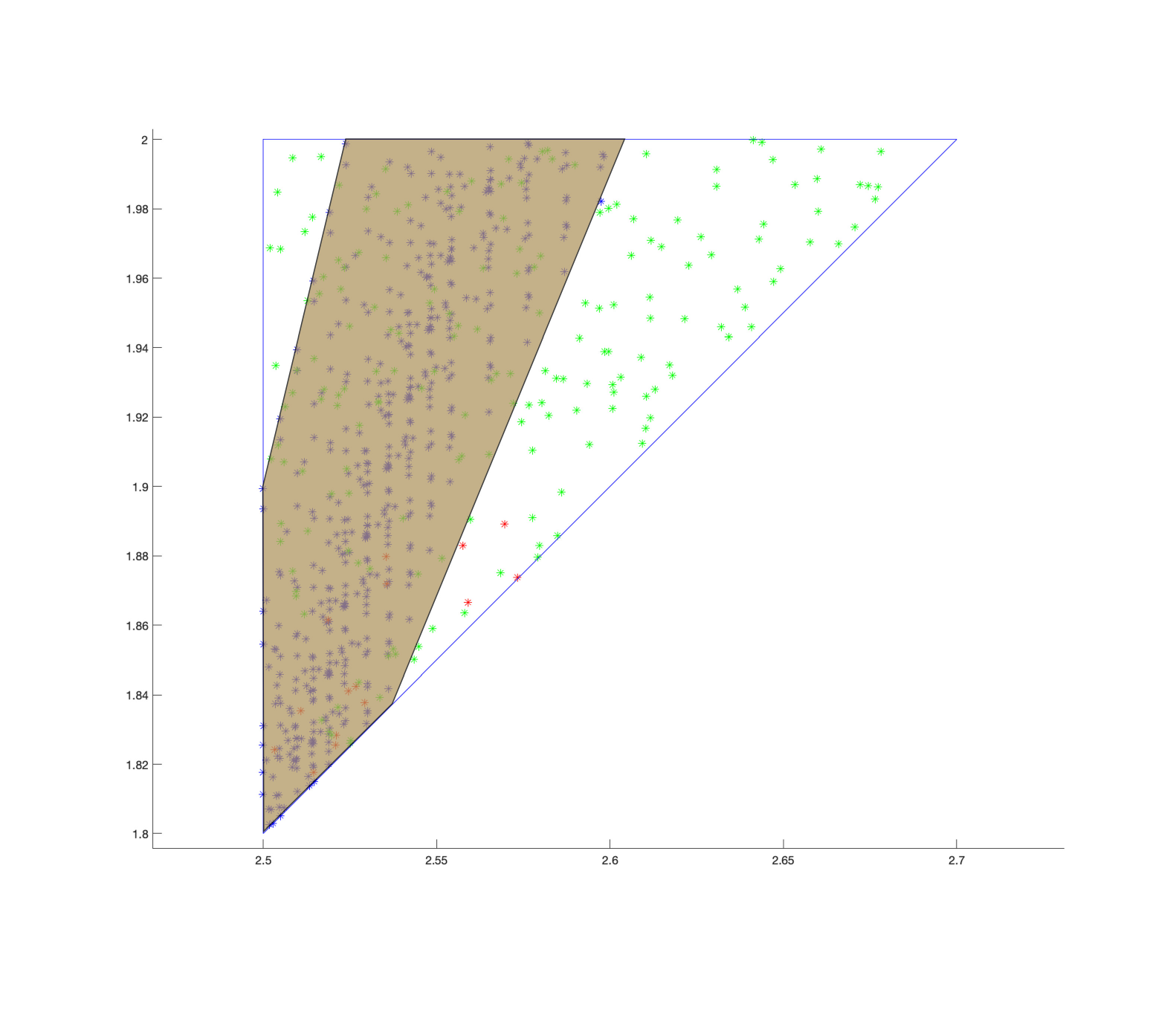}}   
        \end{tabular}
    \end{tabular}
    \caption{{\color{black} The process runs cyclically in the counter-clockwise direction. At the end of each control phase, we mark in green the points in an initial level that are eligible as starting points to go to the next state level if adequate control actions are applied. Meanwhile, the blue points label viable points at a given state level after a phase is completed. The non-viable states are colored in red. For example, consider the Phase-II control actions so that the system goes from a state $X_1$ in the first level to a state $X_2$ in the second level. The intersections of green and blue points represent those points that enable the process to be repeated continually, and we mark this region in yellow color.\label{fig:viability-sim}}} 
\end{figure*}

{\color{black} The plant operating points are depicted in Fig.~\ref{fig:viability-sim}, $\overline{X}_1 = (0,0)$, $\overline{X}_2 = (2.5,2)$, and $\overline{X}_3 = (1,3)$. The operational process is robust and hence it is sufficient for the plant to be in states that are in specified neighborhoods, denoted by $X_1, X_2$, and $X_3$, respectively. The plant process must visit cyclically three states' levels in order, namely, $X_1$, $X_2$, then $X_3$. Operating the plant involves three control phases. In Phase-I (respectively, Phase-II and Phase-III), control inputs are fed to the system in order to reach the state $X_1$ (respectively, $X_2$, $X_3$). To elaborate, let us consider that the plant is operating in Phase-II. The control law of this phase must move the plant state from a first state level (at a point $X_1$) to some point $X_2$ in the second state level. An off-line processing approach is typically used to synthesize state feedback control laws that move the plant state from one operating point to the next \cite{10.1007/3-540-60472-3_7}.}

{\color{black} The plant's simulation results are depicted in Fig.~\ref{a}. The system state and corresponding control input are updated every 0.1, 0.05, and 0.05 time units in Phase-I, Phase-II, and Phase-III, respectively. The green regions depict the viable kernels that are around the ideal states. Starting from any point in a green region, the plant will continually and cyclically run (remains viable) if adequate decision controls are taken. We observed that the TOC policy has a direct impact on the viability of the process. For instance, we experienced a change in the width of the viability kernel (green region) by the simple fact of altering the state/control update frequency, i.e., TOC rate. For low rates, the width of the viability kernel decreases. It is hence crucial to monitor the communication rate a high enough for the proper function of the plant and low enough to save communication resources. Such an example discusses a typical problem that many FoF use cases will raise.}

{\color{black} We design an adaptive TOC update policy that infers an adequate TOC rate according to the position of the state within the admissible region. First, we generate a set of 50 priors for each $X_n$, considering fixed update intervals. We observe the trajectories in each phase and identify the viable states. Suppose that both viable and non-viable priors are within a circle centered in one of the priors states. In such a case, the state is at the edge of the viability kernel. Hence, it is more likely that a trajectory starting from this state will end up non-viable. To enlarge the viability kernel a smaller update interval is scheduled. In Fig.\ref{b}, and \ref{c}, green-colored regions refers to the points in $X_3$ and $X_2$ that allow the system to run swiftly and properly. As shown, by adopting a TOC policy, the viability kernel around $X_2$ has significantly increased, and slightly increased for $X_3$, without increasing the average communication rate. Furthermore, the proposed TOC policy decreased the communication rate by 12\% in Phase-III, and by 14\% in Phase-I.}

{\color{black} Finally, we note that the gain obtained by increasing the communication rate differs with respect to system parameters. In this context, we calculate the transfer entropy between the state and control processes in Table~\ref{tab:TE}. Note that updating the temperature state more often than 0.1 time units does not provide additional valuable information. Thus, a customized policy \textit{per} each system parameter can further improve the TOC efficiency.}

\begin{table}[t]
    \centering
    \caption{Transfer Entropy (bits) between the System States and Control Input during Phase-II. RTransferEntropy package is used to calculate TE from the state and control time-series data.}
    \label{tab:TE}
    \begin{tabular}{|c|c|c|}
    \hline
     \multirow{3}{*}{Update period}&\multicolumn{2}{c|}{Transfer Entropy (state $\rightarrow$ control input)}\\\cline{2-3}
       &\multirow{2}{7em}{Temperature $\rightarrow$ Heat supplied}  & \multirow{2}{8em}{Pressure $\rightarrow$ Piston displacement}  \\&&\\
      \hline
    0.1&0.168&0.014\\
    0.075&0&0.0035\\
    0.05&0&0.0007\\
    \hline
    \end{tabular}

\end{table}

% %---------------------------------------------------------

\vspace{-1mm}
\section{Future Research Directions} % Challenges and Opportunities
\label{sec:challenges}

\subsection{Performance Measurement of TOC} 
\noindent {\bf Experimental viability kernel and TE} --  Deriving the Viability Kernel and quantifying the transfer of knowledge requires a comprehensive understanding of the actor dynamics and the nature of the interactions with the control unit and the environment. It will be challenging, if not possible, to capture all these aspects within a mathematical model. We project that a significant part of the future creative process involves devising methods to measure and create real-like models without operative systems. We recommend relying on a TOC sandbox, much like an AI sandbox. A TOC sandbox is an environment for training, testing, and evaluating AI models before their application to operating networks \cite{9779644}. Further research is required to determine the TOC sandbox's requirements and high-level architecture in future networks. Furthermore, there is a necessity to expand the classical information theory, which traditionally relies on statistical models, to accommodate data-driven models. This involves exploring novel theories and methodologies that can effectively operate within the context of data-driven models.

% \vspace{2mm}
% \noindent %{\bf Need for a universal metric for the effectiveness of communication} --
In this context, \textit{application-agnostic metrics} are way more valuable, providing broader insights and being more universally applicable. They will help design universal TOC algorithms suitable for various environments, goals, and  viability kernels. A trivial metric can be the average size of viability kernels supported by TOC policies. The latter will facilitate an objective comparison of the performance among various TOC policies. An alternative metric could be semantic entropy, which relies on the presence of semantic information representation \cite{DBLP:journals/corr/abs-2201-01389}.

%------------------------------------------------------
%\vspace{-2mm}
\subsection{Re-thinking Network Performance Indicators}

\noindent {\bf Reliability vs.~Robustness} -- Reliability, in its traditional definition, pertains to the amount of successfully delivered data from a source to the destination without any loss or errors. However, as discussed in this work, an actor may still take appropriate actions and evolves through viable states relying on past experiences and the application's context, even in the presence of communications impairment. This puts into question the legacy evaluation metrics of communication systems that are deemed inefficient in capturing the requirements and quality of experience of robust systems with large viable kernels.

In light of the above discussion, it is of interest to analyze the performance of communication systems with respect to the robustness of actors' operations to communication impairments. The latter depends on the end-users' current state, disturbances, and faults, making it a time- and state-dependent metric.

To mathematically define robust communications and understand their performance limits, information, and coding theoretical research are necessary. Much is yet to be discovered in this area, although recent studies of semantic communication, such as \cite{2202.03338}, have shed some light on the topic.

\vspace{2mm}
\noindent {\bf Equality vs.~ Equity} -- 
In communication networks, the resources are normally shared equally among the users, particularly those with the same features. However, we showed through multiple examples that the communications rate varies not only according to the objective but as well as the size of the viability kernel that changes dynamically as the actor evolves from one state to another and while the knowledge is accumulating. In such a case, TOC aims to maintain the sustainability of end-processes; as such, all tasks should remain viable throughout their lifetimes. Equally sharing the resources between similar actors and for the same actor in different states may end up being an unfair and unjustified decision. This traces back to the equality-equity dilemma. From an equity perspective, the resources must be shared to maintain the sustainability of the end-to-end evolution of all the actors. To ensure successful operation, the network should prioritize flows based on the viability kernels of the tasks. For instance, tasks with small viability kernels may require stricter control and frequent state/control updates to avoid failure. It is also important to consider the usefulness of information flow between the plant and the controller, as some states may have a smaller impact on control decisions, which can be measured by transfer entropy.

\subsection{Re-thinking Network Protocols}
\noindent {\bf Sparse communications} -- In the event of scarce spectral resources, opportunistic communication over sparse periods is deemed to be the only feasible strategy to accommodate a large number of actors. A recent breakthrough involves using neural network architectures with a reinforcement learning framework to train multi-agent communication frameworks \cite{9466501}. This includes solving partially observable tasks over noisy channels or common broadcast mediums \cite{kim2019learning}. To ensure agents learn communication protocols that meet their needs, future networks must be flexible. Referring to the example considered earlier, the language model between a controller and plant may require exchanging changes in temperature and pressure. Enforcers, \cite{2201.07452}, can be used to optimize communication within a given budget and ensure sparsity, such as reducing the frequency of state and control updates. Although some research has been conducted on this subject, further exploration is necessary to develop methods for reducing the size of the language and disseminating this knowledge.\\

\newpage

\section{Conclusion}
{\color{black} This article discusses a novel approach for task-oriented communications, utilizing viability theory to enhance communication efficiency. We stressed that the requirements for envisaged 6G applications would be extensive, meanwhile, the spectrum resources remain scarce and costly. By harnessing viability theory, we can devise an end-to-end TOC framework that is robust and productive. We delve into the difficulties and prospects of integrating this theory into task-oriented communications and provide an academic example to showcase its impact on system viability. While our proposed architecture may not be a panacea, we anticipate that it will stimulate further research in this domain.}

\ifCLASSOPTIONcaptionsoff
  \newpage
\fi

\bibliographystyle{IEEEtran}

\bibliography{Learning}

\end{document}